\documentclass[usenatbib,referee]{mn2e}
\usepackage{epsfig}

\begin{document}


\title{The Vela Pulsar's Radio Nebula}

\author[Dodson, Lewis, McConnell, Deshpande]{
  R. Dodson$^1$,
  D. Lewis$^{1,2}$,
  D. McConnell$^2$,
  A. A. Deshpande$^{3,4}$\\
$^1$ School of Mathematics and Physics, University of Tasmania, Hobart
  7000, Australia\\
     Present address: ISAS, Japan.\\
     rdodson@vsop.isas.ac.jp\\
$^2$ Australia Telescope National Facility, CSIRO, PO 76, Epping 1710,
  Australia\\
$^3$ Raman Research Institute, C.V. Raman Avenue, Bangalore 560080, India.\\
$^4$ NAIC/Arecibo Observatory, HC3 Box 53995, Arecibo, Puerto Rico 00612.\\}

\maketitle
\begin{abstract}

We have discovered that the radio nebula surrounding the Vela pulsar
covers a much wider extent than previously reported, with two lobes to
the North and South of the pulsar.
%
Indications of this object have been reported previously, but its
symmetric morphology around the pulsar and other details had not been
identified as they were hidden due to poor sensitivity to low spatial
frequencies.
The structure is highly polarised and the polarisation vectors, once
corrected for Faraday rotation, reveal symmetry with respect to
the spin axis of the pulsar. The X-ray emission found by Chandra lies at the
centre of this structure, in a region which has no detectable excess
of radio emission. 
We estimate total fluxes and regional fluxes from the
Northern and Southern lobes, plus the X-ray region at four radio
frequencies; 1.4, 2.4, 5 and 8.5 GHz.
We present the corresponding images in both the total and
polarised intensities, as well as those showing the derotated 
linear polarisation vectors.

\end{abstract}


\section{Introduction}

The observations of the Vela pulsar wind nebula (PWN) made by Chandra
X-ray Observatory \citep{pavlov_00,helfand} lead us to examine the
published high resolution results of the region
\citep{biet,bock_most,bock_1,bock_2} and the archives of the Australia
Telescope Compact Array (ATCA) to explore whether there was similar
radio emission associated with the pulsar.
Prompted by the revealing images obtained from the available
(archival) data, we re-observed with four moderately compact
configurations, and two closely spaced frequencies. Correlation data
in each integration were folded and binned at the pulsar period
(pulsar binning) at the highest possible rate (thirty two bins per
period).  Furthermore we reprocessed archived data at 1.4 and 2.4-GHz.

Previous high resolution observations of this region are reported in
\citet{biet} using the VLA, \citet{bock_most} using Molonglo
Observatory Synthesis Telescope (MOST) and
\citet{bock_1,bock_2} using the ATCA. These final two papers discuss
results from the same data as we subsequently accessed from the
archive.

Our observations show an extended radio nebula with symmetric
morphology surrounding the pulsar and its X-ray nebula.
We describe our results, compare them to those in the literature, and
consider some of the consequences.

\section{Observations and Data reduction}

All the data were collected at the Australia Telescope Compact Array
(ATCA) in Narrabri (latitude -30.3$^o$) \citep{atca_92}. Observations
were made at 5.2-GHz with the Array in the standard configurations
0.375, 0.75D, 1.5D and 6D. The maximum and minimum baselines were 110
and 0.35 k$\lambda$ (angular resolutions of $1.9^{''}$ to $9.8^{'}$)
for a total of 42.7 hours. Observations at 8.5-GHz we used the
0.75D, 1.5D and EW352 configurations with a four pointing mosaic, to
get full coverage of the extended source. The maximum and minimum
baselines were 130 and 0.6 k$\lambda$ ($1.6^{''}$ to $5.7^{'}$) and
data were collected over a total of 38.3 hours. In both cases we
observed in two nearby frequencies, with bandwidths of 128 MHz, to
ensure that there would be no ambiguities in the rotation measure (RM)
and to maximise the independent {\em uv} coverage.  We used the ATNF
correlator mode that divides each integration data into separate phase
bins spanning the pulsar period.  This firstly allowed the highly
variable pulsar flux
to be excluded from the image and secondly
allowed us to selfcalibrate (i.e. derive the antennae phases from the
observations) on the strong signal from the pulsar (0.6 Jy at
8.5~GHz). Where the interstellar dispersion smeared the pulse arrival
time across the band into several bins (significant only for
observations below 2.4-GHz) we dedispersed across the
multiple spectral channels. The correlator configuration used for our
observations separated the received signal between thirty two pulsar
bins. The archived data from 1996 was observed as part of a thirty
five
pointing mosaic with only one observation covering the pulsar. They
were made with an eight-bin pulsar mode.  The full details are shown
in Table 1.

\begin{table}
\begin{center}
\label{tab:obs}
\begin{tabular}{|l|l|c|c|c|}
\hline
\multicolumn{5}{|c|}{Observations of the Vela Radio Nebula}\\
\hline
\multicolumn{2}{|l|}{Configuration \& date}&Frequency/GHz&Hours&Max and Min baseline/m\\
0.375&01FEB24& 4.85 \& 5.65 & ~8.5 & 5969~~31  \\
0.15D&01MAR17& 4.85 \& 5.65 & ~9.7 & 4439~107   \\
6D   &01MAR30& 4.85 \& 5.65 & 12.1 & 5878~~77  \\
0.75D&01APR18& 4.85 \& 5.65 & 12.4 & 4469~~31  \\
\hline
0.75D&01SEP24& 8.38 \& 8.64 & 11.1 & 4469~~31  \\
EW352&01OCT19& 8.38 \& 8.64 & 11.1 & 4439~~31  \\
1.5D &01OCT29& 8.38 \& 8.64 & ~8.4 & 4439~107  \\
1.5D &01NOV15& 8.38 \& 8.64 & ~7.7 & 4439~107  \\
\hline
0.75C&96JAN09& 1.38 \& 2.37 & ~0.2 & 5020~~46  \\
0.75B&96JAN24& 1.38 \& 2.37 & ~0.2 & 4500~~61  \\
0.75D&96MAY21& 1.38 \& 2.37 & ~0.2 & 4469~~31 \\
6C   &96JUL31& 1.34 \& 1.43 & ~4.2 & 6000~153  \\
0.75A&96NOV22& 1.38 \& 2.37 & ~0.2 & 3750~~77  \\
\hline
\end{tabular}
\vspace{5mm}
\begin{tabular}{|c|l|c|c|c|}
\hline
Frequency&Max Resolution&$\sigma$&Image resolution&$\sigma$\\
(GHz)&(arcsec)&($\mu$Jy per beam)&(arcsec)&($\mu$Jy per beam)\\
1.4&~8.1  $\times$ 6.2&120&8.1  $\times$ 6.2&120\\
2.4&10.5  $\times$ 9.6&270&36   $\times$ 26 &300\\
5.2&~1.8  $\times$ 1.1&30&12.1  $\times$ 10.5&59\\
8.5&~1.5  $\times$ 0.9&74&11.2  $\times$ 10.6&57\\
\hline
\end{tabular}
\vspace{5mm}
\begin{tabular}{|c|c|}
\hline
Frequency&Pointing Centre\\
1.4&08:35:20.67~-45:13:35.79\\
2.4&08:35:20.67~-45:13:35.79\\
5.2&08:35:20.68~-45:10:35.79\\
8.5&08:35:21.88~-45:10:08.80\\
'' &08:35:25.48~-45:11:47.80\\
'' &08:35:15.88~-45:11:47.80\\
'' &08:35:11.08~-45:13:35.80\\
\hline
\end{tabular}
\caption{Summary of the observations}
\end{center}
\end{table}

\begin{figure}
\begin{center}
\epsfig{file=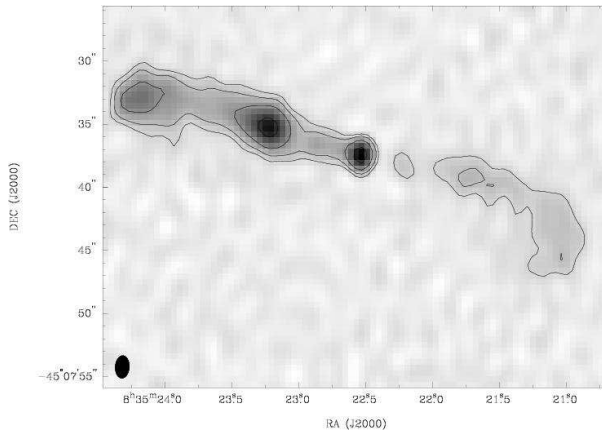, angle=270, width=8cm}
\caption{The background FR-II galaxy at 08:35:22 -45:07:37 from the 5
GHz observations. Contours at 0.125,0.25,0.5,1 mJy/beam. The beam size
is shown in the bottom left. This object is modelled at maximum
resolution and subtracted from {\em uv} in most of the following images.}
\label{fig:fr-ii}
\end{center}
\end{figure}

The data reduction was done following the standard procedures using
the {\small MIRIAD} software package. The antenna gains were
calibrated with regular observations of B0823-500. Absolute flux
calibration was against B1934-638, observed at the end or the
beginning of each observation. The bright FR-II galaxy to the North (centred at
08:35:22.6~-45:07:37.6) has been suppressed by modelling its structure
at the highest resolution and subtracting its contribution 
from the {\em uv} visibilities. Details of this source are to be found
in Table \ref{tab:fr-ii} and the image at 5~GHz in Figure
\ref{fig:fr-ii}. We selfcalibrated on the bin containing the maximum
pulsar flux and transferred the phases to the off-pulse data. All
images were then produced by grading the {\em uv} plane appropriately
for an equivalent FWHM beamwidth of $10^{\prime \prime}$, except the 1.4-GHz and
2.4-GHz images, which were produced with uniform
weighting. Deconvolution was done with the task {\small PMOSMEM}, a
full polarisation version of the maximum entropy task {\small MOSMEM},
for mosaicked images.

\section{Image analysis}

A single pointing was used for the 5~GHz observations and radio
emission was found over a region comparable to the primary beam
($9.3^\prime$).  The images are corrected for the primary beam taper
and are cut off where the response falls below 25\% of the
maximum. The most obvious effect of this is the higher noise levels at
the edge of the images. We have confirmed that we are producing an
accurate representation of this wide field of view using the model
from the mosaicked image in an equivalent {\em uv} data set. As that
model was fully recovered we can be sure that the image produced is
not truncated. The phase centre was $1^{\prime \prime}$ south of the
pulsar.

The next set of observations, at 8.5~GHz, was mosaicked and thus has a
greater sky coverage than those for the 5~GHz observations, despite the
narrower primary beam. These images are corrected for the combined
primary beam taper and have a cutoff at 33\% of the maximum.

A similar, pulsar binned, mosaicked observation of the entire Vela~X
region was made by Bock in 1996 \citep{bock_1,bock_2}. He kindly
made the calibrated 2.4-GHz data available. The 1.4-GHz data from six
128-MHz bands was recovered from the archives with the maximal number
of spectral channels. We processed those pointings of these data that
covered the pulsar in a similar fashion, with the added step of
correction for the pulsar dedispersion across the frequency channels.
In these data we suppressed the FR-II galaxy and
three other point sources (30~mJy at 08:33:43.8~-45:17:58, 13~mJy at
08:34:52.8~-45:20:07 and 7~mJy at 08:35:29.2~-45:18:25) in the field.
The 2.4-GHz data consist of three 128-MHz bands, with poor long
baseline coverage. These data were imaged without the 6~km antenna, so
it has the lowest resolution; $36^{\prime \prime} \times 26^{\prime
\prime}$.

The images for both of the lower frequencies also have corrections for
the primary beam applied. These, however, are much less significant
for the large primary beam at these frequencies. The phase centre for
these observations was $3^\prime$ south of the pulsar.

One of the common problems is in comparing the absolute flux scales
between different configurations responding to different sets of
spatial frequencies.
We have addressed this by making point to point comparisons of images
smoothed to a $20^{\prime \prime}$ beam size, and fitting a global
slope and intercept to this intensity-intensity plot. These showed
that the flux scales agreed with each other to within 1~mJy/beam,
except for the 5-GHz observations, which suffered a shortfall of
3~mJy/beam (i.e. $F_{6cm}=F_{3cm} \times 0.78+3$mJy). Given that the 5
GHz observations were not mosaicked, and therefore could not be
expected to fully recover flux on the shortest baselines, this is not
surprising.
These offsets are corrected for in the spectral index images, and for
the integrated flux density measurements.

\section{Results}

\subsection{The background FR-II galaxy}

We have derived the flux density from the background FR-II galaxy at
each frequency. Low resolution images of the nebula will need to
correct for this, so we provide the fluxes in Table \ref{tab:fr-ii}
and the 5-GHz image (Figure \ref{fig:fr-ii}). The galaxy was imaged
and deconvolved at maximum resolution, and the deconvolution model
subtracted from the {\em uv} data before imaging the region around
the Vela pulsar. 

\begin{table}\begin{center}
\begin{tabular}{|c|c|}
\hline
Frequency&Integrated Flux Density\\
(GHz)&(mJy)\\
\hline
 1.38 & 58 \\ 
 2.37 & 50 \\
 5.23 & 23 \\
 8.51 & ~7 \\  
\hline
\end{tabular}
\end{center}
\caption{Fluxes for FR-II at 08:35:22 -45:07:37}
\label{tab:fr-ii}
\end{table}


\subsection{Images at 8.5~GHz}

\begin{figure}
\begin{center}
\epsfig{file=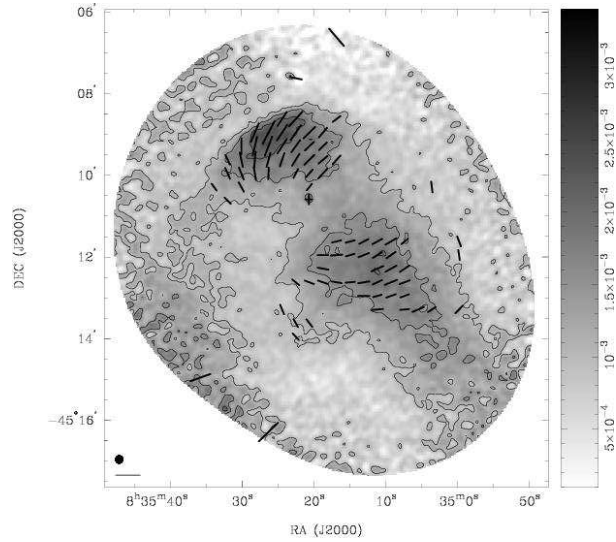, angle=270, width=8cm}
\caption{The Vela radio nebula at 8.5 GHz. The image covers down to
33\% of the primary beam response and the contours are at 1,1.5 and 2
mJy/beam. The image is corrected for this taper and has the FR-II
galaxy removed. The polarisation E-vectors are overlaid, the length of
the bar in the bottom left represents 1~mJy. The beam size is shown in
the bottom left. The pulsar is marked with a cross.}
\label{fig:xband}
\end{center}
\end{figure}

We find a highly polarised and roughly symmetric region of emission
which is centred on the pulsar. The polarised fraction is 30\% with a
r.m.s. variation of 7\% across the nebula. In Figure \ref{fig:xband} the
flux from the pulsar and the FR-II galaxy has been removed.

We have defined the northern lobe as the area with brightness greater
than 1.25~mJy/beam that lies within $2.5^\prime$ north of the pulsar
and $2.5^\prime$ east to $2.5^\prime$ west. This covers an
area of 5.3 square arcminutes in which we find an integrated flux
density of 230~mJy. Altering the cutoff flux by $\pm10$\% gives
variation of 40~mJy. For the more diffuse southern lobe,
we define the region where flux density is greater than 1~mJy, and
extending to $5^\prime$ south of the pulsar, within the band
$1.3^\prime$ east to $5^\prime$ west. This we found to cover 18 square
arcminutes and amount to 650~mJy.
Errors in these estimates arise from uncertainty in the definition of
the regions and the lack of low spatial frequency information.
However, as we have corrected for the offset in the
linear fit to flux against frequency, we believe these uncertainties
are insignificant. The observed flux values, and those from the other
observed bands are listed in Table \ref{tab:flux}.

The flux emitted in the region of the X-ray emission, which we define
as the region where Chandra detected flux greater than 20 $\sigma$
above the background (1 count per square arcsecond), 
is $25\pm3$~mJy. This is a well defined region covering 0.73 square
arcminutes.
Our error limits are those fluxes measured with a X-ray flux cutoff of
$17 \sigma$ and $23 \sigma$. We also estimated the flux from the
region after the subtraction of a planar background. The residual flux
found was within a few sigma of zero. We report both of the total
flux and the residual values in table \ref{tab:flux}.

\subsection{Images at 5~GHz}

\begin{figure}
\begin{center}
\epsfig{file=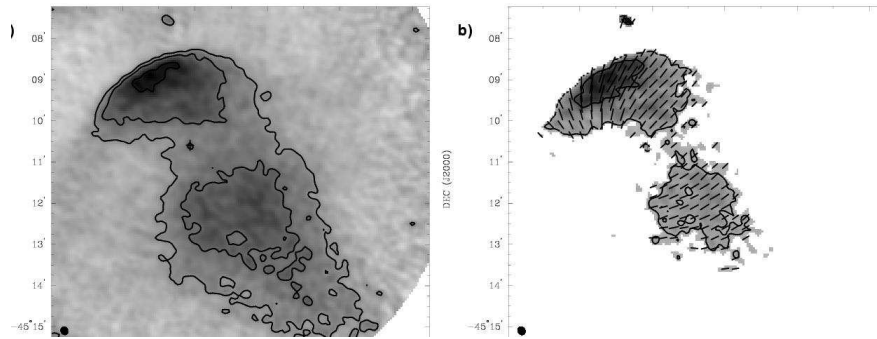, width=12cm}
\caption{a) total intensity at 5~GHz, and b) polarised intensity with
the vectors overlaid. Contours are at 1,2,3 and 4~mJy/beam and the
length of the bar
in bottom left corner represents 1~mJy. The image is corrected for,
and truncated at 25\% of, the primary beam response, and has the FR-II
galaxy removed.}
\label{fig:cband}
\end{center}
\end{figure}

We find an even more highly polarised region of emission at this
frequency. The mean polarised fraction is 60\% with a r.m.s. of 10\%
across the nebula. Figure \ref{fig:cband}a shows the total intensity,
and \ref{fig:cband}b the polarised intensity with position angles
overlaid. We estimated the integrated flux densities by using the
regions defined at 8.5~GHz and offsetting the flux scale as discussed
in section \ref{sec:si}. The results are in Table \ref{tab:flux}.



The region around Vela was observed previously with the VLA
\citep{biet} at 5~GHz, and their image of the region is Figure 1 in
that paper. Their image does not go nearly as deep as ours (stopping
at the $8\sigma$ contour). Therefore they reported only the sharp,
bright, northern edge, and as a result had identified the symmetry
axis (and the spin-axis projection) as that bisecting the length of
the feature. Whilst the observations have similar resolution and point
source sensitivity, our combined configurations have a greater
concentration of the shorter baselines that are most sensitive to the
low surface brightness structures we find here.
%
%
The peak fluxes we find for the FR-II galaxy and the edge of the
northern lobes are consistent with their observations. 



\subsection{Images at 2.4~GHz}

\begin{figure}
\begin{center}
\epsfig{file=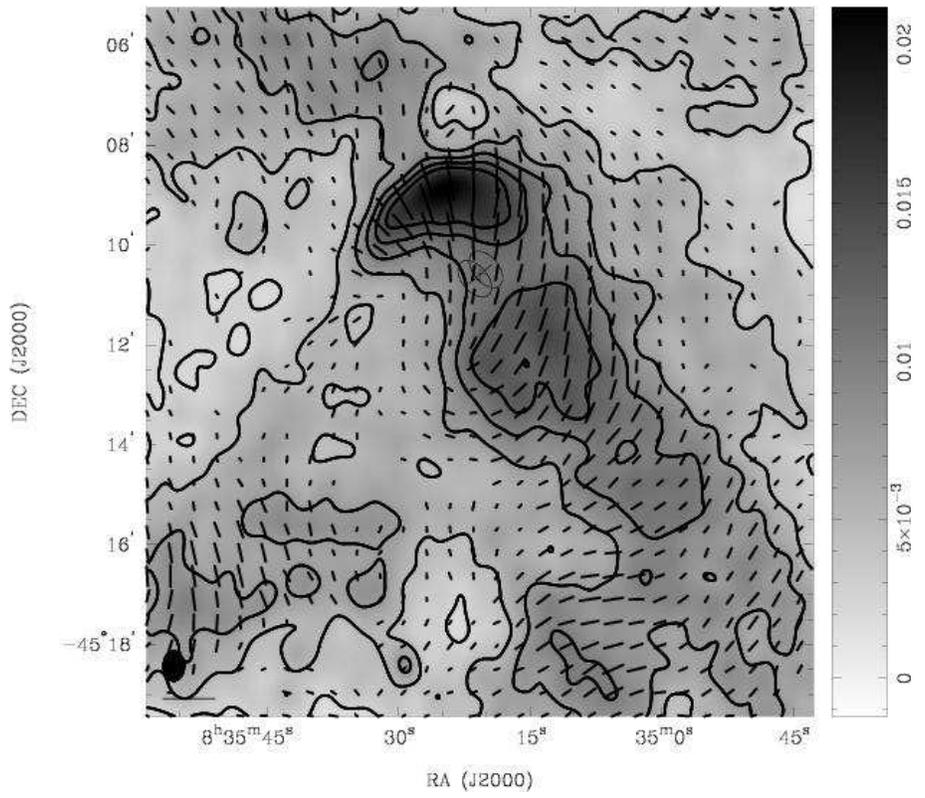, angle=270, width=12cm}
\caption{Total intensity around Vela at 2.4 GHz, with polarisation
E-vectors shown. The length of the bar in the bottom left represents 10~mJy, and the
restoring beam size is shown above it. The image is corrected for
primary beam response and the FR-II galaxy has been removed. A
wire-frame model of the X-ray emission is overlaid. The contour levels
are at 2.5,5,7.5,10 and 12.5 mJy/beam.}
\label{fig:total-s}
\end{center}
\end{figure}

We found the same double lobed structure in these lower resolution
observations. The Southern lobe is more nebulous than at 5~cm, but
clearly does not connect to the bar of emission seen in radio and
X-rays \citep{bar} in the Vela X region (Figure
\ref{fig:total-s}). The fluxes from the three regions defined are
listed in Table \ref{tab:flux}.

\subsection{Images at 1.4~GHz}

\begin{figure}
\begin{center}
\epsfig{file=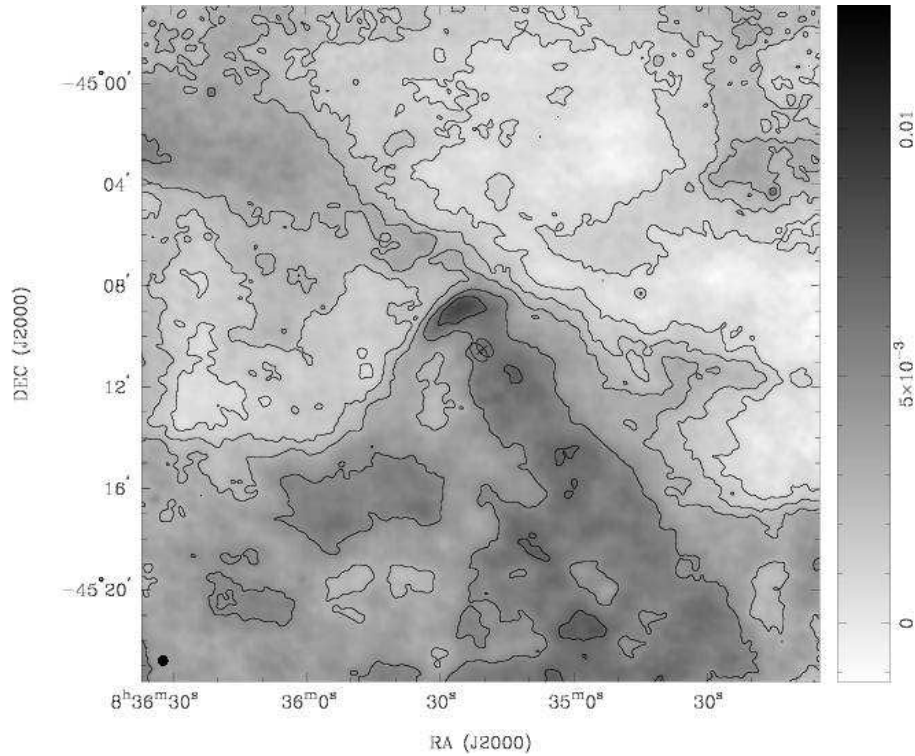, angle=270, width=12cm}
\caption{Total intensity around Vela at 1.4 GHz. Image is corrected
for the primary beam response and the FR-II galaxy has been removed. A
wire-frame model of the X-ray emission is overlaid and the beam size
is shown in the bottom left. There appears to be a ridge of emission
running SW from the pulsar. Contour levels are at 1,2,3,5,7 and 9 mJy/beam}
\label{fig:total-l}
\end{center}
\end{figure}

The data presented here are a subset of the 
observations of the whole Vela SNR region, which are published by
\citet{bock_1,bock_2}. The best of these images were made at
$22^{\prime \prime} \times 23^{\prime \prime}$ resolution. Figure 3 in
\citet{bock_1} covers a slightly larger region than our Figure
\ref{fig:total-l}. Our images were made at the maximum resolution and
then smoothed to match the resolution of the other images ($20^{\prime
\prime} \times 20^{\prime \prime}$), but have the strong background
sources removed. 

Two sources were found (1.1 \& 0.9 mJy) exactly one beam width away to
the North and the South of the pulsar position. They are only found in
the lowest or highest frequency channels of the pulsar bins after or
before the pulsar signal. The pulsar is so strong at these frequencies
that we are seeing the images from the bandpass sidelobes (i.e. about
128~MHz above and below our band edges) which of course have a phase
change of $180^o$. Excluding these the integrated flux from this
region is 33~mJy. The spectral index calculations using the 1.4-GHz
data have these point sources subtracted. The flux from within this
region is 2.8~mJy if we fit and subtract a planar background. 1~mJy of
this is at the position of the pulsar and probably is flux from the
pulse unsuccessfully excluded in the bin selection.

\begin{figure}
\begin{center}
\epsfig{file=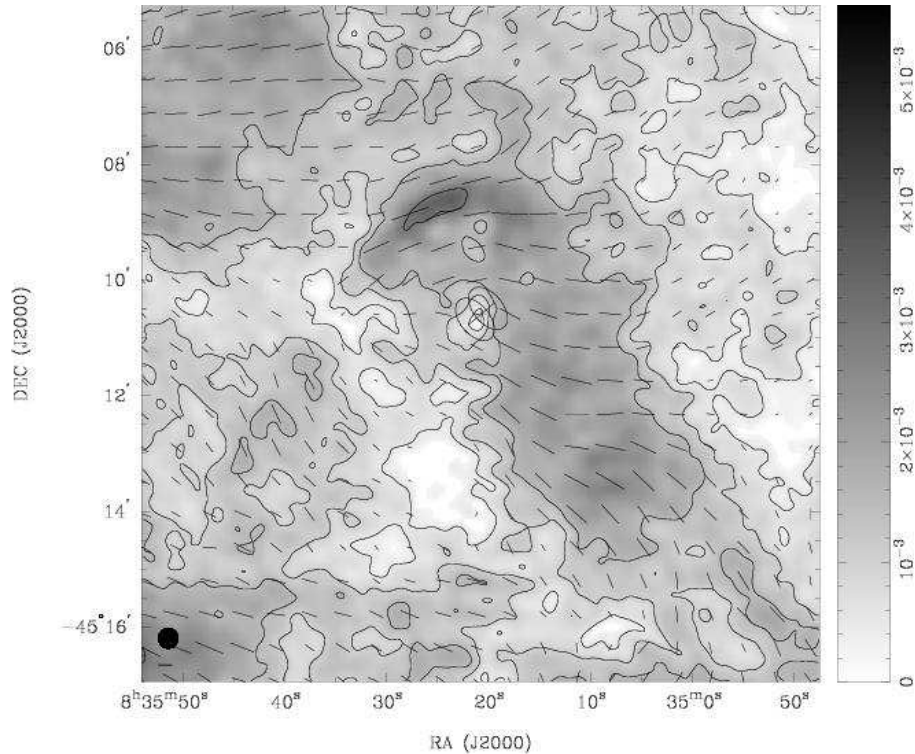, angle=270, width=12cm}
\caption{Polarised intensity and E-vectors around Vela at 1.4 GHz. The
image is corrected for the primary beam response, and the FR-II galaxy has
been removed. The wire-frame model of the X-ray emission is overlaid
and the length of the bar in the bottom left represents 1~mJy. Contour
levels are at 0.5,1,1.5,3 and 4.5 mJy/beam.}
\label{fig:poli}
\end{center}
\end{figure}

These images are complicated by other strong steep spectrum
emission. However the polarised total intensity images, Figure
\ref{fig:poli}, do show a double lobed structure. We have confirmed
that the variation of the polarised emission is not associated with
bandwidth depolarisation by processing sub-bands separately then
recombining them. In particular the measured RM at the edge of the
southern lobe (less than 70~rad~m$^{-2}$) is too small to cause any
significant losses in intensity and thus produce a false edge.

\begin{table}\begin{center}
\begin{tabular}{|c|c|c|c|l|}
\hline
Frequency&\multicolumn{4}{|c|}{Integrated flux density/mJy}\\
GHz&X-ray&less background&North&South\\
\hline
 8.51 & $25\pm3$ &$0.4\pm0.4$& $230\pm40$ & $650\pm70$ \\  
 5.23 & $31\pm4$ &$0.7\pm0.4$& $290\pm50$ & $760\pm100$ \\
 2.37 & $28\pm4$ &$1.1\pm0.6$& $260\pm50$ & $680\pm80$ \\
 1.38 & $33\pm5$ &$2.8\pm0.3$& $260\pm50$ & $820\pm110$ \\ 
\hline
\end{tabular}
\end{center}
\caption{Fluxes for the X-ray region, the Northern and Southern
lobes. The 5-GHz fluxes have an added baseline correction of
3~mJy/beam. The errors are from variation in the definition of the
regions. For the X-ray region this is a $3\sigma$ from the cutoff, for
the others it is 10\%.} 
\label{tab:flux}
\end{table}

\subsection{Spectral Index and rotation measure}
\label{sec:si} 

\begin{figure}
\begin{center}
\epsfig{file=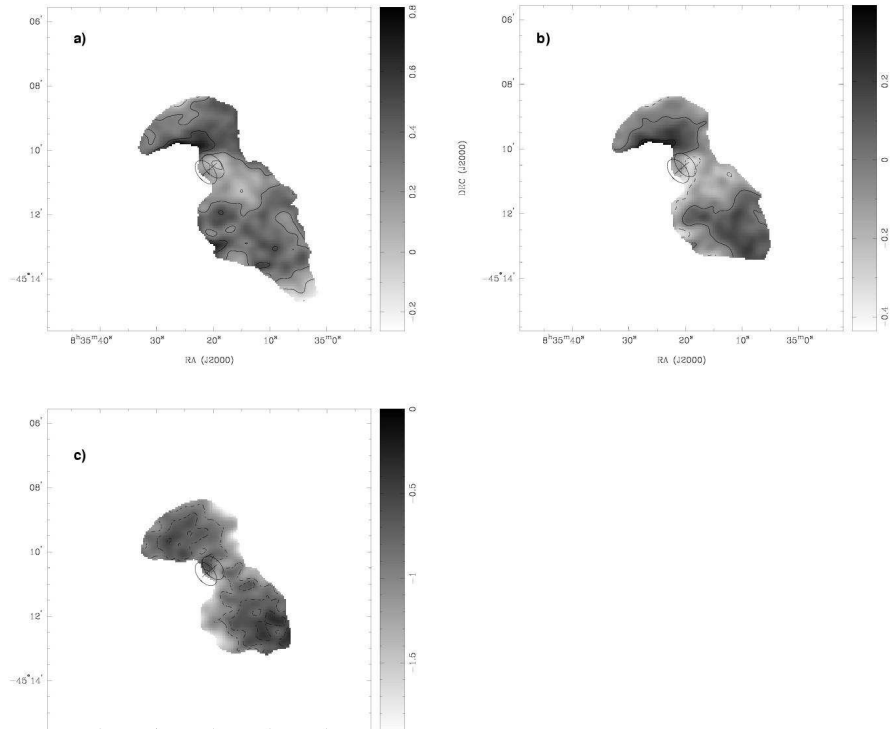, width=12cm}
\caption{Spectral Index between a) 1.4 and 5 GHz, b) 1.4 and 8.5 GHz
and c) 5 and 8.5 GHz. All images have a $20^{\prime\prime}$ resolution
and the flux scale offsets between the observations have been
removed. The model of X-ray emission is overlaid. Contours
levels are overlaid, in steps of 0.25.}
\label{fig:si}
\end{center}
\end{figure}

Comparisons between the images were done after smoothing them to
$20^{\prime \prime}$. The lower resolution spectral index images made
using the 2.4-GHz are not presented. The comparisons between the other
three frequencies are shown in Figure \ref{fig:si}. The mean spectral
index was calculated across the whole region and is listed in Table
\ref{tab:si}. The errors are the one sigma variation across the image,
not that of each independent point. The spectral index was calculated
after comparing
the data at different frequencies in an `intensity-intensity' plot, to
find the offset in the flux scales between the frequencies. This
avoids the perennial problem in comparing interferometric images with
different {\em uv} sampling. It was found that the dominant offset was
for 5-GHz which was underestimated by 3~mJy/beam (at $20^{\prime\prime}$).
The formal errors are dominated by the error in the flux scale
offset, and are not uniform. They are typically $\pm0.03$ in regions
with significant flux density.

\begin{table}\begin{center}
\begin{tabular}{|c|c|c}
\hline
Frequencies (GHz)&$\alpha$&$\sigma_\alpha$\\
\hline
 1.38/5.23&~0.3&0.1\\ 
 1.38/8.51&~0.0&0.1\\
 5.23/8.51&-0.9&0.3\\
\hline
\end{tabular}
\end{center}
\caption{Spectral Index between 1.4, 5 and 8.5 GHz. Errors are the
variation across the source.}
\label{tab:si}
\end{table}

In most cases the spectral index is uncorrelated with known features,
except for between 5-GHz and 1.4-GHz where $\alpha = 0.3\pm0.1$ (where
$F\propto\nu^{\alpha}$) across the image, but with a noticeable
flattening around the X-ray region. Further observations to
investigate this are required. The edges of the lobes in the 5/8-GHz
spectral index are remarkably high, but have the poorest signal to
noise.

\begin{figure}
\begin{center}
\epsfig{file=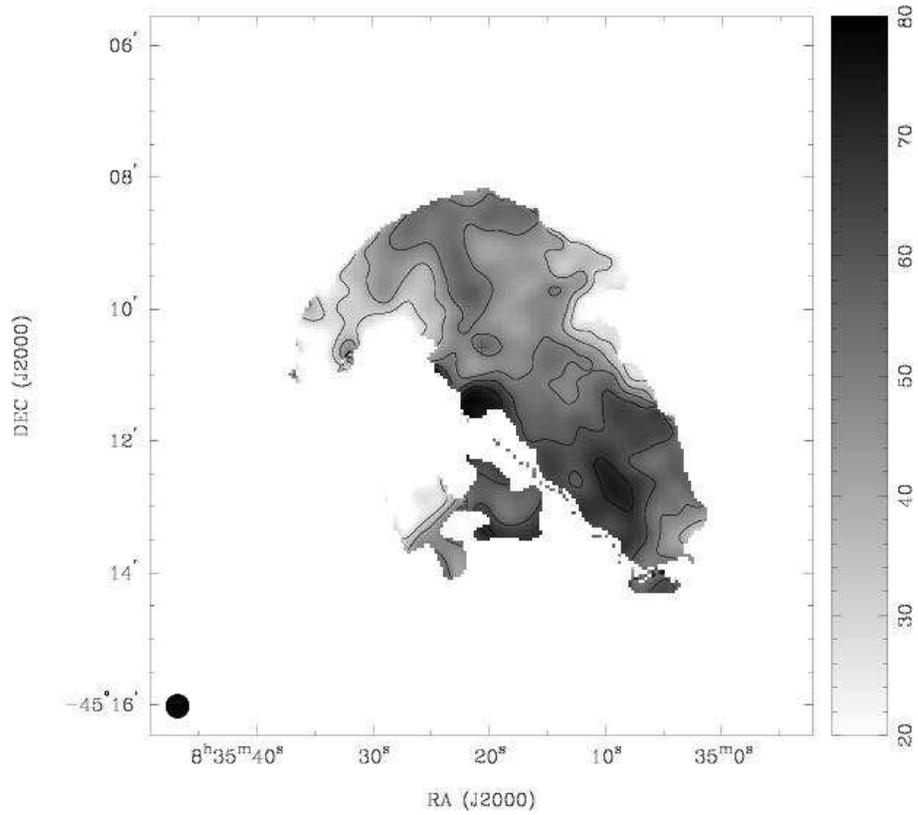, angle=270, width=12cm}
\caption{The rotation measure across the source, found between 2.4, 5
and 8.5 GHz. The cross marks the pulsar position. Contours are at RM of
30,40,50,60,70 rads m$^{-2}$}
\label{fig:rm}
\end{center}
\end{figure}

Faraday rotation along the sight line from the nebula has been deduced
from the multi-frequency observations.  Figure \ref{fig:rm} shows its
variation across the source. After correction for this propagation
effect, the intrinsic orientation of the linearly polarised component
(Figure \ref{fig:pa0}) shows a symmetric pattern about the pulsar spin
axis projection.

The above rotation measure values imply that there would be some
bandwidth depolarisation (a differential Faraday rotation of $56^o$)
across the 20~cm band. Averaging across the band, however,
would only produce a 5\% fall in the integrated polarised flux. This
effect could be dealt with, but would add little to our results, and
has not been attempted.




\subsection{Polarised intensity}


Polarisation fraction and angle provides one of the greatest
diagnostics as to the underlying processes and conditions. We see in
the 5 and 8-GHz observations that the polarisation fraction is high,
and approaches the theoretical maximum for synchrotron emission. 
The polarisation position angles (associated with the electric field
vectors) are first corrected for Faraday rotation, then are rotated
further by 90 degrees to view (as shown in Figure \ref{fig:pa0}) the
corresponding magnetic field structure in the emission regions. A
clear signature of toroidal magnetic field with respect to the spin
axis of the pulsar is apparent.


A more detailed model of the emission, combining the X-ray and radio
data, will be presented in the next paper.

\subsection{Pulsar Wind Nebula surface brightness}

Using the latest estimated distance of 0.3~kpc
\citep{caraveo_vela_pm,dodson_vela_pm} and using equation 4 from
\citet{frail_97} we find the radio luminosity to be $6.8\times10^{30}$
ergs/s, or $1\times10^{-6}$ of the spin down energy, $\dot{E}$. This
is below the upper limit for twenty seven of the thirty one
non-detections of PWN discussed in \citet{gaensler_00}. 
It may be that the lack of PWN is not
due to their intrinsic rarity, but their inefficiency in converting
the spin down power into detectable radio emission. Furthermore the
assumption of a proper motion powered shock front clearly breaks down
as the bright radio emission features are approximately parallel to
the proper motion direction and there is no `wake-like' feature that
can be associated with the known proper motion direction (which is
close to the projected spin axis of the star).


\section{Conclusions}

We believe we have observed the true extent and morphology of the
radio wind nebula associated with the Vela pulsar.  The toroidal
magnetic field structure revealed by the polarisation data and the
symmetric morphology around the projected spin axis of the star (as in
the X-ray case) argue strongly for the nebula to be pulsar (rotation)
driven, in terms of both, the recent and continued injection of particles
and the (possibly wound-up) magnetic fields in the regions of radio
emission. The extension towards the south, seen in the 2.4 and 1.4 GHz
images, is unpolarised and thus has a different character from that of
the nebula. The rotation measure along these sight-lines appears
insufficient to cause significant depolarisation.



This allows the much more complex region, Vela X, to be
unambiguously identified with the integrated history of Vela's
emission, but not driven by the current particle population. This
affects all the estimates of expected PWN brightness based on the
Vela-X region (e.g. \citet{frail_97}). 


The most striking feature in comparing the X-ray and radio emissions
is that the radio emission starts at a distance from the centre where
X-ray emission stops. Given the remarkable similarity in the
morphologies in the two bands, it is tempting to model the radio
nebula on the lines of the available models for the X-ray data.
Results of our preliminary attempts \citep{lewis_02} to extend the
model of \citet{desh_rad} for the radio data have been very
encouraging. A more detailed modelling of the radio data is in
progress and will be reported later separately.





\section{Acknowledgements}

The Australia Telescope Compact Array is part of the Australia
Telescope, funded by the Commonwealth of Australia for operation as a
National Facility, and managed by CSIRO. The access of data from the
archive was assisted by Robin Wark. Dr Bock 
made the calibrated 2.4-GHz data available. The pulsar binning mode
required the timing observations made at Mount Pleasant. This research
has made use of NASA's Astrophysics Data System Abstract Service. The
X-ray data for the Chandra observations were downloaded from the
public access site, the Chandra Data Archive (CDA). This is part of
the Chandra X-Ray Observatory Science Centre (CXC) which is operated
for NASA by the Smithsonian Astrophysical Observatory.

The 1.4~GHz data was partially processed by Godfrey and See-Toh for
their 3rd-year astronomy practical. 


\begin{figure}
\begin{center}
\epsfig{file=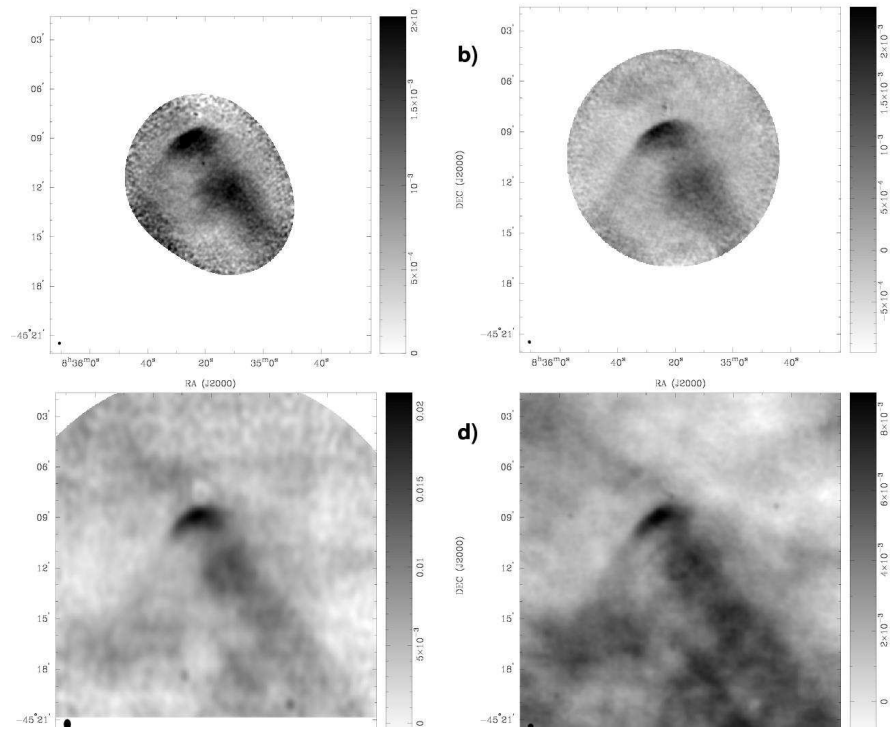, width=12cm}
\caption{The Vela radio nebula at 8.5 (top left), 5.2 (top right), 2.4
(bottom left) and 1.4-GHz (bottom right) on the same scale.}
\label{fig:all}
\end{center}
\end{figure}

\begin{figure}
\begin{center}
\epsfig{file=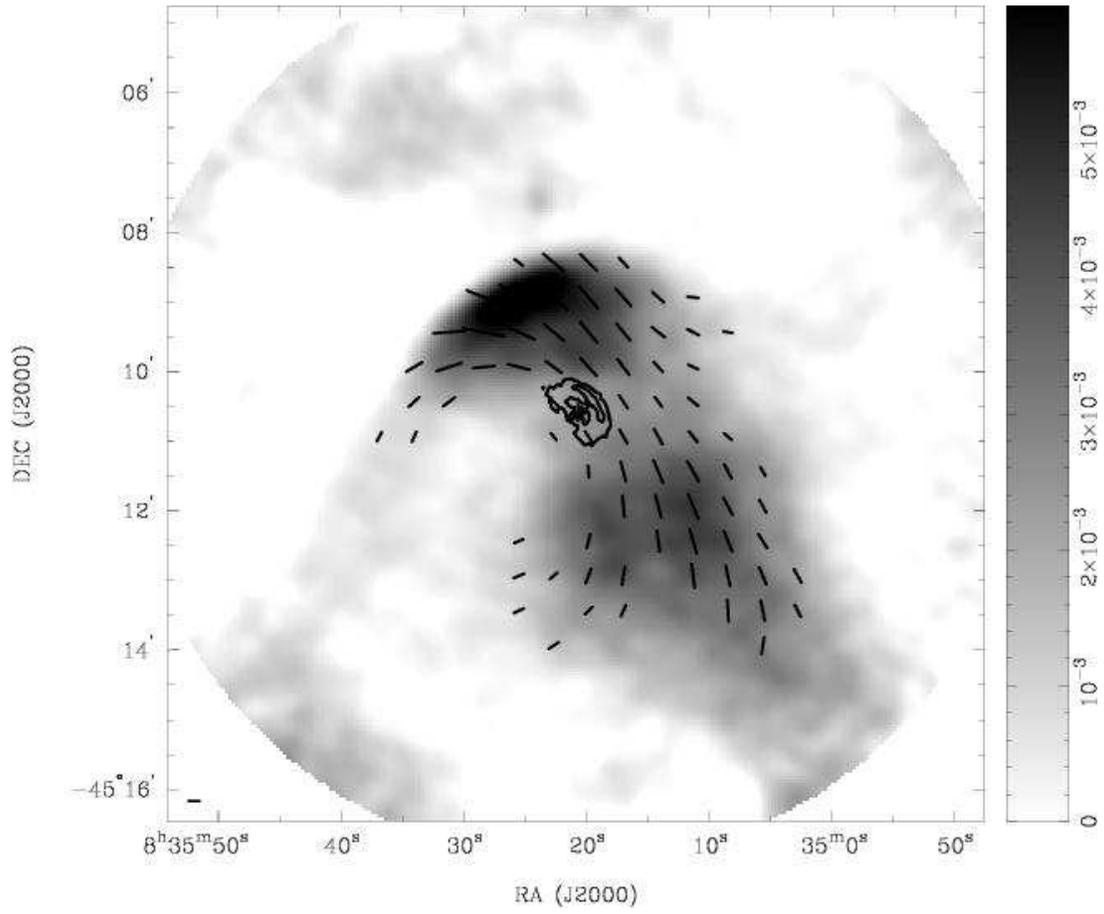, width=12cm, angle=270}
\caption{Vela PWN at 5~GHz with derotated magnetic field lines
overlaid. and the length of the bar in the bottom left represents
1~mJy. The contours from the Chandra observations, at 30 and 70
$\sigma$, are overlaid.}
\label{fig:pa0}
\end{center}
\end{figure}

\end{document}